\documentclass{article}
\usepackage{amssymb}
\usepackage{amsmath}

\begin{document}
\centerline{\Large \bf The general solutions of some nonlinear
second order PDEs.}
\medskip
\centerline{\Large \bf I. Two independent variables, constant
parameters}

\vskip 2cm \centerline{\sc Yu. N. Kosovtsov}
\medskip
\centerline{Lviv Radio Engineering Research Institute, Ukraine}
\centerline{email: {\tt kosovtsov@escort.lviv.net}} \vskip 1cm
\begin{abstract}
In the first part of planned series of papers the formal general
solutions to selection of 80 examples of different types  of second
order nonlinear PDEs in two independent variables with constant
parameters are given. The main goal here is to show on examples the
types of solvable PDEs and what their general solutions look like.
The solving strategy, used here, as a rule is the order reduction.
The order reduction method is implemented in Maple procedure, which
applicable to PDEs of different order with different number of
independent variables. Some of given PDEs are solved by order
lifting to PDEs, which are  solvable by the subsequent order
reduction.
\end{abstract}

\section{Introduction}

Nonlinear partial differential equations (PDEs) play very important
role in many fields of mathematics, physics, chemistry, and biology,
and numerous applications. Despite the fact that various methods for
solving nonlinear PDEs have been developed in 19-20 centuries \cite
{Darb1}-\cite {Zhiber}, there exists a very disadvantageous opinion
that only a very small minority of nonlinear second- and
higher-order PDEs admit general solutions in closed form (see, e.g.,
percentage of PDEs with general solutions in fundamental handbook
\cite {Pol}).

Nevertheless there exist some extensive nontrivial families for
different types of nonlinear PDEs which general solutions can be
expressed in closed form and which seemingly are not described in
literature.

In the first part of planned series of papers the formal general
solutions to selection of 80 examples of different types  of second
order nonlinear PDEs in two independent variables with constant
parameters are given. The main goal here is to show on examples the
types of solvable PDEs and what their general solutions look like.

The solving strategy, used here, as a rule is the order reduction.
The order reduction method is implemented in Maple procedure (see
Appendix), which applicable to PDEs of different order with
different number of independent variables \cite{Kos}. Some of given
PDEs are solved by order lifting to PDEs, which are  solvable by the
subsequent order reduction.

I try to to follow the tradition when solvable PDEs are filed up to
point transformations, deciding between equivalent PDEs variants the
most short one.

I would like to thank Prof. A.D. Polyanin for valuable advices in
results presentation.

\section{Equations of the Form {\mathversion{bold}{$\displaystyle \frac{\partial^2 w}{\partial
t\partial x}= F(w,\frac{\partial w}{\partial t},\frac{\partial
w}{\partial x}) $}}}
\vskip 0.5cm


\subsection {\mathversion{bold}{$\ \displaystyle \frac{\partial^2 w}{\partial
t\partial x}=\frac{2}{w}\left[\frac{\partial w}{\partial
t}+1\right]\frac{\partial w}{\partial x}\,.$}} \vskip 0.5cm

\vskip 0.5cm General solution

\begin{equation}
\displaystyle w(x,t) =-\frac{F(t)}{F'(t)}+
\left\{\left[F'(t)\right]^2\left[G(x)-\int\frac{F''(t)}{F(t)\left[F'(t)\right]^2}\,dt\right]
\right\}^{-1}\,, \notag
\end{equation}
where $F(t)$ and $G(x)$ are arbitrary functions.

\vskip 0.5cm


\subsection {\mathversion{bold}{$ \displaystyle \left( \frac{\partial^2
w}{\partial t \, \partial x} \right)^2 = \left( \frac{\partial
w}{\partial x} \right)^2 \frac{\partial w}{\partial t}\,.$}} \vskip
0.5cm

General solution

\begin{equation}
\displaystyle  w(t,x) = -\frac{4F'(t)}{F(t)+G(x)}+\int
\left(\frac{F''(t)}{F'(t)}\right)^2\,dt\,, \notag
\end{equation}
where $F(t)$ and $G(x)$ are arbitrary functions.\vskip 0.5cm


\subsection {\mathversion{bold}{$\displaystyle \left(\frac{\partial^2 w}{\partial
t\partial x}-w\frac{\partial w}{\partial
x}\right)^2+\left(2\frac{\partial w}{\partial
t}-w^2\right)\left(\frac{\partial w}{\partial x}\right)^2=0\,.$}}
\vskip 0.5cm

 \vskip 0.5cm General
solution

\begin{equation}
\displaystyle w(t,x) =
\left(-\frac{1}{2}\int(F'(t))^2\,e^{F(t)}\,dt+G(x)\right)e^{-F(t)}\,,
\notag
\end{equation}
where $F(t)$ and $G(x)$ are arbitrary functions. \vskip 0.5cm


\subsection {\mathversion{bold}{$\displaystyle \left(\frac{\partial w}{\partial
t}+w\right)^2 +\left(\frac{\partial^2 w}{\partial t\partial
x}+\frac{\partial w}{\partial x}+ \frac{\partial w}{\partial
t}+w\right)\times\\\\\\\exp \left(\frac{\frac{\partial^2 w}{\partial
t\partial x}+\frac{\partial w}{\partial x}+ \frac{\partial
w}{\partial t}+w}{\frac{\partial w}{\partial t}+w}\right)=0\,.$}}
\vskip 0.5cm

 \vskip 0.5cm General
solution

\begin{equation}
\displaystyle w(t,x) = \left(e^{-x}\int e^t\,F(t)\exp
[F(t)\,e^{-x}]\,dt+G(x)\right)e^{-t}\,, \notag
\end{equation}
where $F(t)$ and $G(x)$ are arbitrary functions. \vskip 0.5cm


\subsection {\mathversion{bold}{$\displaystyle \frac{\partial^2 w}{\partial
t\partial x}= a\left(\frac{\partial w}{\partial
t}+w\right)^n-\frac{\partial w}{\partial x}\,,$}} \vskip 0.5cm

where $a\neq 0$, and $n\neq 1$ are constants.\vskip 0.5cm General
solution

\begin{equation}
\displaystyle w(t,x) = \left(a^{\frac{1}{1-n}}\int
e^t\left[F(t)+(1-n)x\right]^{\frac{1}{1-n}}\,dt+G(x)\right)e^{-t}\,,
\notag
\end{equation}
where $F(t)$ and $G(x)$ are arbitrary functions. \vskip 0.5cm


\subsection {\mathversion{bold}{$\displaystyle \frac{\partial^2 w}{\partial
t\partial x} -\frac 1w\frac{\partial w}{\partial t}\frac{\partial
w}{\partial x}-\frac{c}{w}\frac{\partial w}{\partial t}-k w= 0\,,$}}
\vskip 0.5cm

where $c,k$ are constants. \vskip 0.5cm General solution

\begin{equation}
\displaystyle w(t,x) = \frac{\{\displaystyle
-c\int\exp[x(1-kt)]G(x)\,dx+F(t)\}\exp[-x(1-kt)]}{G(x)}\,, \notag
\end{equation}
where $F(t)$ and $G(x)$ are arbitrary functions.\vskip 0.5cm


\subsection {\mathversion{bold}{$\displaystyle \frac{\partial^2w}{\partial
t\partial x}  = \frac{1}{w}\left(\frac{\partial w}{\partial x}
+a\right )\frac{\partial w}{\partial t}+bw\frac{\partial w}{\partial
x}\,,$}} \vskip 0.5cm

where $a$ and $b\neq 0$ are constants. \vskip 0.5cm General solution

\begin{equation}
\displaystyle w(t,x) = -\frac{1}{b}\frac{\displaystyle
\int_{-\infty}^\infty\frac{1}{\omega}\left[F(\omega)ab\exp
\left(\frac{abt+\omega^2x}{\omega}\right )+G(\omega)\omega^2\exp
\left(\frac{abx+\omega^2t}{\omega}\right )\right
]\,d\omega}{\displaystyle \int_{-\infty}^\infty\left[F(\omega)\exp
\left(\frac{abt+\omega^2x}{\omega}\right )+G(\omega)\exp
\left(\frac{abx+\omega^2t}{\omega}\right )\right ]\,d\omega}\,,
\notag
\end{equation}
where $F(\omega)$ and $G(\omega)$ are arbitrary functions. \vskip
0.5cm


\subsection {\mathversion{bold}{$\displaystyle \frac{\partial^2 w}{\partial
t\partial x}- \left(\frac 1w\frac{\partial w}{\partial
t}+b\right)\frac{\partial w}{\partial x}-\frac{c}{w}\frac{\partial
w}{\partial t}-c b= 0\,,$}} \vskip 0.5cm

where $b,c$ are constants. \vskip 0.5cm General solution

\begin{equation}
\displaystyle w(t,x) =
\left\{-c\int\,\exp[-e^{bt}G(x)]\,dx+F(t)\right\}\exp[e^{bt}G(x)]\,,
\notag
\end{equation}
where $F(t)$ and $G(x)$ are arbitrary functions.\vskip 0.5cm


\subsection {\mathversion{bold}{$\displaystyle \frac{\partial^2 w}{\partial
t\partial x}= \left[\frac{(2w-a-1)}{(w-1)(w-a)}\frac{\partial
w}{\partial x}+\frac{b(w-a)}{w-1}\right]\frac{\partial w}{\partial
t}\,,$}} \vskip 0.5cm

where $a\neq 1$, and $b$ are constants. \vskip 0.5cm General
solution

\begin{equation}
\displaystyle w(t,x) =
\frac{G'(x)+abG(x)+aF(t)}{G'(x)+bG(x)+F(t)}\,, \notag
\end{equation}
where $F(t)$ and $G(x)$ are arbitrary functions. \vskip 0.5cm


\subsection {\mathversion{bold}{$\displaystyle \frac{\partial^2 w}{\partial
t\partial x}= \frac{(2kw-ak-c)}{(w-a)(kw-c)}\frac{\partial
w}{\partial t}\frac{\partial w}{\partial
x}+\frac{2b(kw-c)^2}{ak-c}\,,$}} \vskip 0.5cm

where $a,b,c$, and $k$ are constants. \vskip 0.5cm General solution

\begin{equation}
\displaystyle w(t,x) =
\frac{aF'(t)G'(x)-c[F(t)-bG(x)]^2}{F'(t)G'(x)-k[F(t)-bG(x)]^2}\,,
\notag
\end{equation}
where $F(t)$ and $G(x)$ are arbitrary functions. \vskip 0.5cm


\subsection {\mathversion{bold}{$\displaystyle \left(\frac{\partial^2 w}{\partial t\partial
x}+\frac{\partial w}{\partial x}\right)^2
+\\\\\\a^2\left[\left(\frac{\partial w}{\partial
t}+w\right)^2-b^2\right] \left(\frac{\partial w}{\partial
t}+w\right)^2=0\,,$}} \vskip 0.5cm

where $a\neq 0$, and $b\neq 0$ are constants. \vskip 0.5cm General
solution

\begin{equation}
\displaystyle w(t,x) =
\left(2b\int\frac{F(t)\exp(abx+t)}{1+F^2(t)\exp(2abx)}\,dt+G(x)\right)e^{-t}\,,
\notag
\end{equation}
where $F(t)$ and $G(x)$ are arbitrary functions. \vskip 0.5cm


\subsection {\mathversion{bold}{$\displaystyle \frac{\partial^2 w}{\partial t\partial
x}+\frac{\partial w}{\partial x} +a\left[\left(\frac{\partial
w}{\partial t}+w\right)^2+b^2\right] \left(\frac{\partial
w}{\partial t}+w\right)=0\,,$}} \vskip 0.5cm

where $a\neq 0$, and $b\neq 0$ are constants. \vskip 0.5cm General
solution

\begin{equation}
\displaystyle w(t,x) = \left(\pm
b\int\frac{e^t\,dt}{\sqrt{F(t)\exp[(2ab^2x)-1}}+G(x)\right)e^{-t}\,,
\notag
\end{equation}
where $F(t)$ and $G(x)$ are arbitrary functions. \vskip 0.5cm


\subsection {\mathversion{bold}{$\displaystyle \left(w\frac{\partial^2 w}{\partial
t\partial x}-\frac{\partial w}{\partial t}\frac{\partial w}{\partial
x}\right)\exp\left[\frac{a\frac{\partial^2 w}{\partial t\partial
x}}{\frac{\partial w}{\partial x}}\right]=b\frac{\partial
w}{\partial x} \,,$}} where $a$, and $b$ are constants.
 \vskip 0.5cm General
solution

\begin{equation}
\displaystyle w(t,x) = \left(-b\int
\exp\left[aF'(t)+F(t)\right]\,dt+G(x)\right)e^{-F(t)}\,, \notag
\end{equation}
where $F(t)$ and $G(x)$ are arbitrary functions. \vskip 0.5cm


\subsection {\mathversion{bold}{$\displaystyle \left(\frac{w\frac{\partial^2 w}{\partial t\partial
x}}{\frac{\partial w}{\partial x}}- \frac{\partial w}{\partial t}
\right)^2=w \left(a\frac{\partial w}{\partial x}-bw\right)\,,$}}
where $a\neq 0$, and $b$ are constants.
 \vskip 0.5cm General
solution

\begin{align}
\displaystyle w (t,x) =
F(t)\exp\left\{\frac{b}{a}\int\left(cos\left[ \frac
{t\sqrt{b}}{2}+G(x) \right] \right)^{-2}\,dx\right\} \,, \notag
\end{align}
where $F(t)$ and $G(x)$ are arbitrary functions. \vskip 0.5cm


\subsection {\mathversion{bold}{$\displaystyle \frac{\partial^2 w}{\partial t\partial
x}+b\frac{\partial w}{\partial x}=a\exp\left[- \frac{\partial
w}{\partial t}-bw\right]\,,$}} where $a$ and $b$ are constants.
 \vskip 0.5cm General
solution

\begin{align}
\displaystyle w (t,x) =
\left(\int\,\ln\left[ax+F(t)\right]e^{bt}\,dt+G(x)\right)e^{-bt} \,,
\notag
\end{align}
where $F(t)$ and $G(x)$ are arbitrary functions. \vskip 0.5cm


\subsection {\mathversion{bold}{$\ \displaystyle(aw+1)\left(\frac{\partial w}{\partial
t}+b\right)\frac{\partial^2 w}{\partial t\partial
x}=\\\\\\a\frac{\partial w}{\partial x}\left(\frac{\partial
w}{\partial t}+b\right)^2-(aw+1)^3\,,$}} \vskip 0.5cm

where $a$, $b$ are constants. \vskip 0.5cm General solution

\begin{align}
\displaystyle w(x,t) = -&\left\{
\int\left(b\pm\sqrt{F(t)-2x}\right)\exp\left[\pm
a\int\sqrt{F(t)-2x}\,
 dt\right]dt+G(x)\right\}\times\notag\\\notag\\\notag
 &\exp\left[\mp a\int\sqrt{F(t)-2x}\,
 dt\right]\,, \notag
\end{align}
where $F(t)$ and $G(x)$ are arbitrary functions.

\vskip 0.5cm


\subsection {\mathversion{bold}{$\displaystyle \left[\left(\frac{\partial w}{\partial
t}\right)^2-2aw \right] \left(w\frac{\partial^2 w}{\partial
t\partial x}-\frac{\partial w}{\partial t}\frac{\partial w}{\partial
x}\right)^2=\\\\ \left[w\frac{\partial^2 w}{\partial t\partial
x}\frac{\partial w}{\partial t}+aw\frac{\partial w}{\partial
x}-\left(\frac{\partial w}{\partial t}\right)^2\frac{\partial
w}{\partial x} \right]^2\,,$}} where $a\neq 0$ is a constant.

 \vskip 0.5cm General
solution

\begin{equation}
\displaystyle w(t,x) =
\left(-\frac{a}{2}\int\frac{e^{F(t)}}{F'(t)}\,dt+G(x)\right)e^{-F(t)}\,,
\notag
\end{equation}
where $F(t)$ and $G(x)$ are arbitrary functions. \vskip 0.5cm


\subsection {\mathversion{bold}{$\displaystyle \frac{\partial^2 w}{\partial t\partial
x}=\frac{w+1}{w}\frac{\partial w}{\partial t}\frac{\partial
w}{\partial x}+aw^{(1-m)}e^{w(1-m)}\,,$}} where $a$, and $m\neq 1$
are constants.
 \vskip 0.5cm General
solution  in implicit form
\begin{equation}
\displaystyle \int_1^\infty \frac{e^{-\xi
w(t,x)}}{\xi}\,d\xi=G(x)-\int\left[a(1-m)x+F(t)\right]^{\frac{1}{1-m}}\,dt\,,
\notag
\end{equation}
where  $F(t)$ and $G(x)$ are arbitrary functions. \vskip 0.5cm


\subsection {\mathversion{bold}{$\displaystyle w\left(\frac{\partial^2 w}{\partial t\partial
x}+a \frac{\partial w}{\partial t} \right)^2-\frac{\partial
w}{\partial t}\left(\frac{\partial^2 w}{\partial t\partial x}+a
\frac{\partial w}{\partial t}
\right)\times\\\\\\\left(2aw+\frac{\partial w}{\partial
x}\right)+b\left(2aw+\frac{\partial w}{\partial x}\right)^2=0\,,$}}
where $a$ and $b$ are constants.
 \vskip 0.5cm General
solution

\begin{align}
\displaystyle w (t,x) = \left( b \int \frac {\exp \left[-ax-e^{ax}F
( t ) \right]}{F' ( t )}\,dt +G ( x )  \right) \exp \left[e^{ax}F( t
) \right] \,, \notag
\end{align}
where $F(t)$ and $G(x)$ are arbitrary functions. \vskip 0.5cm


\subsection {\mathversion{bold}{$\displaystyle \left(b^2-4aw \frac{\partial w}{\partial
x}\right) \left(aw^2\frac{\partial^2 w}{\partial t\partial
x}-aw\frac{\partial w}{\partial t}\frac{\partial w}{\partial x}+b^2
 \frac{\partial w}{\partial x} \right)^2=\\\\ b^2\left(aw^2\frac{\partial^2 w}{\partial t\partial
x}-3aw\frac{\partial w}{\partial t}\frac{\partial w}{\partial x}+b^2
 \frac{\partial w}{\partial x} \right)^2\,,$}}
where $a$, and $b\neq 0$ are constants.

 \vskip 0.5cm General
solution

\begin{equation}
\displaystyle w(t,x) =
\pm\left(b\int\sqrt{F'(t)}\,e^{aF(t)}\,dt+G(x)\right)e^{-aF(t)}\,,
\notag
\end{equation}
where $F(t)$ and $G(x)$ are arbitrary functions. \vskip 0.5cm


\subsection {\mathversion{bold}{$\displaystyle w\frac{\partial^2 w}{\partial t\partial
x}=\left(\frac{\partial w}{\partial x}-aw\right)\frac{\partial
w}{\partial t}-b\frac{\partial w}{\partial x}-cw^2+abw\,,$}} where
$a\neq 0$, $b\neq 0$, and  $c$ are constants.
 \vskip 0.5cm General
solution
\begin{equation}
\displaystyle w(t,x) =
\left(b\int\exp\left\{\frac{ct}{a}+e^{-ax}F(t)\right\}
\,dt+G(x)\right)\exp\left\{-\frac{ct}{a}-e^{-ax}F(t)\right\}\,,
\notag
\end{equation}
where $F(t)$ and $G(x)$ are arbitrary functions. \vskip 0.5cm


\subsection {\mathversion{bold}{$\displaystyle \frac{\partial^2 w}{\partial
t\partial x}= \left(\frac 1w\frac{\partial w}{\partial
t}+b\right)\frac{\partial w}{\partial x}+\frac{c}{w}\frac{\partial
w}{\partial t}+kw+cb\,,$}} \vskip 0.5cm

where $b\neq0$, $c$, and $k\neq0$ are constants. \vskip 0.5cm
General solution

\begin{equation}
\displaystyle w(t,x) =
\left\{-c\int\exp\left(\frac{k}{b^2}[e^{bt}G(x)+bx]\right)dx+F(t)\right\}\exp\left\{-\frac{k}{b^2}[e^{bt}G(x)+bx]\right\}\,,
\notag
\end{equation}
where $F(t)$ and $G(x)$ are arbitrary functions. \vskip 0.5cm


\subsection {\mathversion{bold}{$\displaystyle \frac{\partial^2 w}{\partial
t\partial x}= \frac{a}{w}\left(\frac{\partial w}{\partial
x}\right)^{\!2}+\frac 1w\frac{\partial w}{\partial t}\frac{\partial
w}{\partial x} +\left(b+\frac{c}{w}\right)\frac{\partial w}{\partial
x}+\\\\\\ \frac{c}{2aw}\frac{\partial w}{\partial
t}+\frac{(bw+c)^2}{4aw}\,,$}} \vskip 0.5cm

where $a\not=0$ and $b,c$ are constants. \vskip 0.5cm General
solution

\begin{equation}
w(t,x) =
\left\{-\frac{c}{2a}\int\exp\left[\frac{1}{2a}\int\frac{2\,dx}{t+G(x)}+bx\right]dx+F(t)\right\}
\exp\left[-\frac{1}{2a}\int\frac{2\,dx}{t+G(x)}+bx\right]\,, \notag
\end{equation}
where $F(t)$ and $G(x)$ are arbitrary functions. \vskip 0.5cm


\subsection {\mathversion{bold}{$\displaystyle
\frac{a^2}{w^4}\left(c\frac{\partial w}{\partial t}+\frac{\partial
w}{\partial t}\frac{\partial w}{\partial x}-w\frac{\partial^2
w}{\partial t\partial x}\right)^{\!2}=\\\\\\
\frac{ac}{w}+b+\frac{a}{w}\frac{\partial w}{\partial x}\,,$}} \vskip
0.5cm

where $a\neq0$, $b$, and $c$ are constants. \vskip 0.5cm General
solution

\begin{align}
\displaystyle w(t,x) =& \left\{-c\int
\exp\left[-\frac{1}{4a}\left(-4bx+\int(t+G(x))^2dx\right)\right]\,dx+F(t)\right\}\times\notag
\\ \notag \\&\exp\left[\frac{1}{4a}\left(-4bx+\int(t+G(x))^2dx\right)\right]\,,
\notag
\end{align}
where $F(t)$ and $G(x)$ are arbitrary functions.\vskip 0.5cm


\subsection {\mathversion{bold}{$\displaystyle w(\frac{\partial w}{\partial x}+aw+b)\frac{\partial^2 w}{\partial
t\partial x}-\frac{\partial w}{\partial t}\frac{\partial w}{\partial
x}(\frac{\partial w}{\partial x}+a w+2b)+\\\\\\b(a
w+b)\frac{\partial w}{\partial t}+c w^3\,,$}} where $a, b,c$ are
constants.

\vskip 0.5cm General solution

\begin{align}
\displaystyle w(t,x)=&
-\left\{b\int\exp\left[-\int\left(\pm\sqrt{G(x)+2ct}-a\right)dx\right]dx+F(t)\right\}\times\notag\\\notag\\\notag
&\exp\left[\int\left(\pm\sqrt{G(x)+2ct}-a\right)dx\right] \,, \notag
\end{align}
where $F(t)$ and $G(x)$ are arbitrary functions. \vskip 0.5cm


\subsection {\mathversion{bold}{$\displaystyle \frac{\partial^2 w}{\partial
t\partial x}-\frac{a}{w}\left(\frac{\partial w}{\partial
x}\right)^{\!2}-\left(\frac{1}{w}\frac{\partial w}{\partial
t}+b+\frac{c}{w}\right)\frac{\partial w}{\partial x}
-\frac{c}{2aw}\frac{\partial w}{\partial
t}-kw-\frac{bc}{2a}-\frac{c^2}{4aw} = 0\,,$}} \vskip 0.5cm

where $a\neq0$, $b$, $c$, and $k$ are constants. \vskip 0.5cm
General solution for $b^2-4ka\neq0$

\begin{align}
\displaystyle &w(t,x) =\notag\\\notag\\&
-\frac{c}{2a}\left\{\int\exp\left[\frac{1}{2a}\int\frac{\exp(t\sqrt{b^2-4ak})G(x)(b+\sqrt{b^2-4ak})-\sqrt{b^2-4ak}+b}{1+\exp(t\sqrt{b^2-4ak})G(x)}dx\right]dx
+F(t)\right\} \times \notag\\\notag\\& \times
\exp\left[-\frac{1}{2a}\int\frac{\exp(t\sqrt{b^2-4ak})G(x)(b+\sqrt{b^2-4ak})-\sqrt{b^2-4ak}+b}{1+\exp(t\sqrt{b^2-4ak})G(x)}dx\right]
\,, \notag
\end{align}
where $F(t)$ and $G(x)$ are arbitrary functions. \vskip 0.5cm


\subsection {\mathversion{bold}{$\ \displaystyle \frac{\partial^2 w}{\partial
x\partial t}-\left(\frac{1}{w}\frac{\partial w}{\partial t}+b\right)
\frac{\partial w}{\partial x}-\frac{c}{w}\frac{\partial w}{\partial
t}- aw^2\left(c+kw+\frac{\partial w}{\partial
x}\right)^{\!-1}\\\\\\-sw-b c = 0  \,,$}} \vskip 0.5cm

where $a,b,c,k,s$ are constants. \vskip 0.5cm General solution

\begin{equation}
w(x,t) = e^{-kx}\exp\left[\int W(x,t)\,dx\right]\left\{-c\int
e^{kx}\exp\left[-\int W(x,t)\,dx\right]\,dx+F(t)\right\}\,, \notag
\end{equation}
where $F(t)$ and $G(x)$ are arbitrary functions, and the $W=W(x,t)$
is any root of the transcendental equation
\begin{equation}
\displaystyle\int^W_0\frac{\xi\,d\xi}{(s-
bk)\xi+b\xi^2+a}=t+G(x).\notag
\end{equation}

\vskip 0.5cm


\subsection {\mathversion{bold}{$\displaystyle (aw^n+b)\frac{\partial^2 w}{\partial t\partial
x}-k(aw^n+b)^{(2-m)} \left(-\frac{\partial w}{\partial
t}\right)^m=\\\\\\\left[anw^{(n-1)}\frac{\partial w}{\partial
x}-c(aw^n+b)\right]\frac{\partial w}{\partial t}\,,$}} where $a$,
$b$, $c$, $k$, $n$, and $m\neq 1$ are constants.
 \vskip 0.5cm General solution  in
implicit form ($w=w(t,x)$)

\begin{equation}
\displaystyle
\int^w_s\frac{d\xi}{a\xi^n+b}+\int\left[F(t)e^{c(m-1)x}-\frac{k}{c}\right]^{\frac{1}{1-m}}\,dt+G(x)=0
\,, \notag
\end{equation}
where $F(t)$ and $G(x)$ are arbitrary functions, $s$ is a constant.
\vskip 0.5cm


\subsection {\mathversion{bold}{$\displaystyle \frac{\partial^2 w}{\partial t\partial
x}=\frac{w-n}{w}\frac{\partial w}{\partial t}\frac{\partial
w}{\partial x}+aw^{n(m-1)}e^{w(1-m)}\,,$}} where $a\neq 0$, $n\neq
-1$, and  $m\neq 1$ are constants.
 \vskip 0.5cm General
solution  in implicit form
\begin{align}
\displaystyle
&w(t,x)^{\frac{n}{2}}\exp\left[-\frac{w(t,x)}{2}\right]M_{\frac{n}{2},\frac{n+1}{2}}\left(w(t,x)\right)=\\\notag \\
&
(n+1)[a(1-m)]^{\frac{1}{1-m}}\int[x+F(t)]^{\frac{1}{1-m}}\,dt+G(x)\,,
\notag
\end{align}
where $M_{p,q}(z)$ is the Whittaker M function, $F(t)$ and $G(x)$
are arbitrary functions. \vskip 0.5cm


\subsection {\mathversion{bold}{$\ \displaystyle (aw^2+bw+cb-ak)\frac{\partial^2 w}{\partial
t\partial x}=\\\\\\\left[(2aw+b)\frac{\partial w}{\partial
t}+w^2+2cw+k\right]\frac{\partial w}{\partial x}\,,$}} \vskip 0.5cm

where $a, b,c$, and $k$ are constants. \vskip 0.5cm General solution

\begin{equation}
\displaystyle w(x,t) =F(t)-W(t)
\left\{G(x)+\int\frac{W(t)\left[aF'(t)+F(t)+c
\right]}{aF^2(t)+bF(t)-ak+cb}\,dt \right\}^{-1}\,, \notag
\end{equation}
where $F(t)$ and $G(x)$ are arbitrary functions, and
\begin{equation}
\displaystyle
W(t)=\exp\left\{\int\frac{\left[2aF(t)+b\right]F'(t)+F^2(t)+2cF(t)+k}{aF^2(t)+bF(t)-ak+bc}\,dt
\right\}.\notag
\end{equation}
\vskip 0.5cm


\subsection {\mathversion{bold}{$\ \displaystyle(aw+b)\frac{\partial^2 w}{\partial t\partial
x}=\left(\frac{\partial w}{\partial t}
\right)^2+\left(a\frac{\partial w}{\partial x}+2cw+2k
\right)\frac{\partial w}{\partial t}+\\\\\\(ak-bc)\frac{\partial
w}{\partial x}+(cw+k)^2\,,$}} \vskip 0.5cm

where $a,b,c$, and $k$ are constants. \vskip 0.5cm General solution

\begin{align}
\displaystyle w(x,t) = &\left\{
-k^2\int\frac{F(t)+x}{k[F(t)+x]-b}\exp\left[\int\frac{ck[F(t)+x]+ak-bc}{k[F(t)+x]-b}\,
 dt\right]dt+G(x)\right\}\times\notag\\\notag\\\notag
 &\exp\left[-\int\frac{ck[F(t)+x]+ak-bc}{k[F(t)+x]-b}\,
 dt\right]\,
  \notag
\end{align}
where $F(t)$ and $G(x)$ are arbitrary functions.

\vskip 0.5cm


\subsection {\mathversion{bold}{$\ \displaystyle w(aw-b)\left(\frac{\partial^2 w}{\partial
t\partial x}\right)^2=(2aw-b)\frac{\partial w}{\partial
t}\frac{\partial w}{\partial x}\frac{\partial^2 w}{\partial
t\partial x}+\\\\\\c(2aw-b)^2\left(\frac{\partial w}{\partial
x}\right)^2\,,$}} \vskip 0.5cm

where $a, b$, and $c$ are constants. \vskip 0.5cm General solution

\begin{align}
\displaystyle&w(x,t) =F(t)+W(t)\times \notag\\\notag\\\notag
&\left\{G(x)-a\int\frac{W(t)\left[\left(F'(t)\right)^2+F'(t)H(t)-2bcF(t)+2acF^2(t)
\right]}{F(t)\left[aF(t)-b\right]\left[F'(t)+H(t) \right]}\,dt
\right\}^{-1}\,, \notag
\end{align}
where $F(t)$ and $G(x)$ are arbitrary functions, and
\begin{equation}
\displaystyle
W(t)=\exp\left\{\int\frac{\left[2aF(t)-b\right]\left[\left(F'(t)\right)^2+F'(t)H(t)-2bcF(t)+2acF^2(t)
\right]}{F(t)\left[aF(t)-b\right]\left[F'(t)+H(t) \right]}\,dt
\right\},\notag
\end{equation}

\begin{equation}
\displaystyle
H(t)=\pm\sqrt{\left(F'(t)\right)^2+4acF^2(t)-4bcF(t)}.\notag
\end{equation}
\vskip 0.5cm


\subsection {\mathversion{bold}{$\displaystyle \frac{\partial^2 w}{\partial t\partial
x}+akw\exp\left[\frac{1}{aw}\frac{\partial w}{\partial
t}+\frac{c}{aw}+b\right]-\\\\\\\frac{1}{w}\left[\frac{\partial
w}{\partial t}+c\right]\frac{\partial w}{\partial x}+m\frac{\partial
w}{\partial t}+abmw+mc=0 \,,$}} where $a\neq 0$, $b$, $c$, $k$ and
$m$ are constants.

\vskip 0.5cm General solution

\begin{align}
\displaystyle w(t,x) =&\left\{-c\int\,\exp\left[abt-a\int W(t,x)\,dt
\right]\,dt+G(x)\right\}\times\notag\\\notag\\\notag
 &\exp \left[-abt+a\int W(t,x)\,dt \right]\,,
\notag
\end{align}
here $W(t,x)$ is any solution of the following transcendental
equation

\begin{equation}
\displaystyle \int_s^{W(t,x)}\frac{d\xi}{ke^\xi+m\xi}+x+F(t)=0\,,
\notag
\end{equation}
where $F(t)$ and $G(x)$ are arbitrary functions, $s$ is any
constant. \vskip 0.5cm


\subsection {\mathversion{bold}{$\displaystyle \left(\frac{\partial^2 w}{\partial
t\partial x}\right)^2+2a\frac{\partial^2 w}{\partial t\partial
x}\frac{\partial w}{\partial x}+b\left(\frac{\partial w}{\partial
t}\right)^3+\\\\(3abw-c^2)\left(\frac{\partial w}{\partial
t}\right)^2+aw(3abw-2c^2)\frac{\partial w}{\partial
t}+\\\\a^2\left(\frac{\partial w}{\partial
x}\right)^2+a^2w^2(abw-c^2)=0\,,$}} where $a$, and $b\neq 0$, $c\neq
0$ are constants.
 \vskip 0.5cm General
solution

\begin{equation}
\displaystyle w(t,x) =
\left(\frac{4c^3e^{cx}}{b}\int\frac{F(t)[F(t)-1]e^{at}}{\left[F(t)(c+e^{cx})-c\right]^2}
\,dt+G(x)\right)e^{-at}\,, \notag
\end{equation}
where $F(t)$ and $G(x)$ are arbitrary functions. \vskip 0.5cm


\subsection {\mathversion{bold}{$\displaystyle w\left(\frac{\partial^2 w}{\partial t\partial
x}\right)^2-\frac{\partial w}{\partial x}\left(\frac{\partial
w}{\partial t}-kw+b\right)\frac{\partial^2 w}{\partial t\partial
x}-\\\\\\\left(\frac{\partial w}{\partial
x}\right)^2\left(k\frac{\partial w}{\partial
t}-a+bk\right)+c\left[\left(\frac{\partial w}{\partial
t}+kw\right)^2+\right.\\\\\\\left.2b\frac{\partial w}{\partial
t}-2w(2a-bk)+b^2\right]\left(\frac{\partial w}{\partial
x}-cw\right)=0\,,$}} where $a$, $b$, $c$, and $k$ are constants.
 \vskip 0.5cm General
solution

\begin{align}
\displaystyle w(t,x) =
\left[-\int\left(\frac{ae^{cx}}{F'(t)}+b\right)\exp\left[kt+\right.\right.&\left.\left.F(t)e^{-cx}\right]\,dt
+G(x)\right]\times\notag \\\notag \\&
\exp\left[-kt-F(t)e^{-cx}\right]\,, \notag
\end{align}
where $F(t)$ and $G(x)$ are arbitrary functions. \vskip 0.5cm


\subsection {\mathversion{bold}{$\displaystyle \left(k\frac{\partial^2 w}{\partial t\partial
x}+a\frac{\partial w}{\partial
x}\right)^2\times\\\\\\\left[\left(k\frac{\partial w}{\partial t}+aw
\right)^2-2bm\frac{\partial w}{\partial t}-2bcw
 \right]=\\\\\\\left[\frac{\partial^2 w}{\partial t\partial
x}\left(k^2\frac{\partial w}{\partial
t}+akw-bm\right)+\right.\\\\\\\left.ak\frac{\partial w}{\partial
t}\frac{\partial w}{\partial x}+\frac{\partial w}{\partial
x}(a^2w-cb)\right]^2\,,$}} where $a$, $b$, $c$, $k$, and $m$ are
constants.
 \vskip 0.5cm General
solution

\begin{equation}
\displaystyle w(t,x) =-\frac{b}{2(am-ck)F(t)}
\left(\int\frac{\left[cF(t)-mF'(t)\right]^2}{aF(t)-kF'(t)}
\,dt+G(x)\right)\,, \notag
\end{equation}
where $F(t)$ and $G(x)$ are arbitrary functions. \vskip 0.5cm


\subsection {\mathversion{bold}{$\displaystyle \frac{\partial^2 w}{\partial t\partial
x}=-\frac{a}{w^2}\left(\frac{\partial w}{\partial
t}\right)^3-\left(\frac{b}{w}+\frac{c}{w^2}\right)\left(\frac{\partial
w}{\partial t}\right)^2+\\\\\\\left[\frac{1}{w}\frac{\partial
w}{\partial
x}+\frac{(ak-b)(3ak+b)}{4a}-\frac{2bc}{3aw}-\frac{c^2}{3aw^2}
\right]\frac{\partial w}{\partial
t}+\\\\\\\frac{c}{3aw}\frac{\partial w}{\partial
x}-\frac{k(ak-b^2)w}{4a}+\frac{c(ak-b)(3ak+b)}{12a^2}-\\\\\\\frac{bc^2}{9a^2w}-\frac{c^3}{27a^2w^2}\,,$}}
where $a\neq 0$, $b$, $c$, and  $k$ are constants.
 \vskip 0.5cm General
solution
\begin{align}
\displaystyle w(t,x) =&\frac{1}{3a}
\left(-c\int\exp\left\{-\frac{1}{2a}\int\frac{(ak-b)V(t,x)-2ak}{V(t,x)+1}\,dt
\right\}\,dt+G(x)\right)\times\notag \\\notag \\
& \exp\left\{\frac{1}{2a}\int\frac{(ak-b)V(t,x)-2ak}{V(t,x)+1}\,dt
\right\}\,, \notag
\end{align}
here
\begin{equation}
\displaystyle V(t,x)=W\left(F(t)\exp\left\{-\frac{(3ak-b)^2x+4a}{4a}
\right\}\right) \,, \notag
\end{equation}
where $W(z)$ is the Lambert W function, $F(t)$ and $G(x)$ are
arbitrary functions. \vskip 0.5cm


\subsection {\mathversion{bold}{$\ \displaystyle bw\frac{\partial^2 w}{\partial
t\partial x}=a\left(\frac{\partial w}{\partial
t}\right)^2+\left(b\frac{\partial w}{\partial
x}+cw+2af\right)\frac{\partial w}{\partial t}+\\\\\\bf\frac{\partial
w}{\partial x}+\frac{c^2g(1-g)}{a}w^2+cfw+af^2\,,$}} \vskip 0.5cm

where $a\neq 0, b\neq 0, c\neq 0$, and $g, f$ are constants. \vskip
0.5cm General solution

\begin{align}
\displaystyle&w(x,t) =-\frac{1}{c}+\frac{W(t,x)}{a}\times
\notag\\\notag\\\notag &\left \{G(x)+
\int\frac{\left[a(af+g-1)\exp\left(\frac{c}{b}[x+2gF(t)]
\right)-(af-g)\exp\left(\frac{c}{b}[2gx+F(t)] \right)
\right]}{W(t,x)\left[-a\exp\left(\frac{c}{b}[x+2gF(t)]
\right)+\exp\left(\frac{c}{b}[2gx+F(t)] \right) \right]} \,dt
\right\}\,, \notag
\end{align}
where $F(t)$ and $G(x)$ are arbitrary functions, and
\begin{equation}
\displaystyle
W(t,x)=\exp\left\{\frac{c}{a}\int\frac{\left[a(g-1)\exp\left(\frac{c}{b}[x+2gF(t)]
\right)+g\exp\left(\frac{c}{b}[2gx+F(t)] \right)
\right]}{\left[a\exp\left(\frac{c}{b}[x+2gF(t)]
\right)-\exp\left(\frac{c}{b}[2gx+F(t)] \right) \right]}
\,dt\right\}.\notag
\end{equation}

\vskip 0.5cm


\subsection {\mathversion{bold}{$\ \displaystyle (a^2bw^4+ck^2)\left(\frac{\partial^2 w}{\partial
t\partial x}\right)^2-2a^2w^3\frac{\partial w}{\partial
x}\left(2b\frac{\partial w}{\partial t}+k\right)\frac{\partial^2
w}{\partial t\partial x}\\\\\\+4a^2kw^2\left(\frac{\partial
w}{\partial x}\right)^2\frac{\partial w}{\partial
t}+4a^2bw^2\left(\frac{\partial w}{\partial
t}\right)^2\left(\frac{\partial w}{\partial x}\right)^2=0\,,$}}
\vskip 0.5cm

where $a, b,c$, and $k$ are constants. \vskip 0.5cm General solution
\begin{equation}
\displaystyle w(x,t) =F(t)-W(t)
\left\{G(x)+a\int\frac{W(t)F'(t)\left[aF^2(t)+H(t)\right]}{a^2F^4(t)+aF^2(t)H(t)-2ckF'(t)}\,dt
\right\}^{-1}\,, \notag
\end{equation}
where $F(t)$ and $G(x)$ are arbitrary functions, and
\begin{equation}
\displaystyle
W(t)=\exp\left\{2a\int\frac{F(t)F'(t)\left[aF^2(t)+H(t)\right]}{a^2F^4(t)+aF^2(t)H(t)-2ckF'(t)}\,dt
\right\},\notag
\end{equation}

\begin{equation}
\displaystyle
H(t)=\pm\sqrt{a^2F^4(t)-4cb\left[F'(t)\right]^2-4ckF'(t)}.\notag
\end{equation}

\vskip 0.5cm


\subsection {\mathversion{bold}{$\ \displaystyle (aw^6+b)\left(\frac{\partial^2 w}{\partial
t\partial x}\right)^3-\\\\\\2w^5\frac{\partial w}{\partial
x}\left(3a\frac{\partial w}{\partial
t}+c\right)\left(\frac{\partial^2 w}{\partial t\partial
x}\right)^2\\\\\\+4w^4\left(\frac{\partial w}{\partial
x}\right)^2\frac{\partial w}{\partial t}\left(3a\frac{\partial
w}{\partial t}+2c\right)\frac{\partial^2 w}{\partial t\partial
x}-\\\\\\8w^3\left(\frac{\partial w}{\partial
x}\right)^3\left(\frac{\partial w}{\partial
t}\right)^2\left(a\frac{\partial w}{\partial t}+c\right)=0\,,$}}
\vskip 0.5cm

where $a$, and $ b\neq 0, c\neq 0$ are constants. \vskip 0.5cm
General solution
\begin{align}
\displaystyle &w(x,t) =F(t)-W(t)\times\notag\\\notag\\\notag &
\left\{G(x)+\int\frac{W(t)F'(t)\left[12bF'(t)F^2(t)+c^3P^2(t)\right]}{12bF'(t)F^4(t)-6bcF'(t)P(t)+c^3F^2(t)P^2(t)}\,dt
\right\}^{-1}\,, \notag
\end{align}
where $F(t)$ and $G(x)$ are arbitrary functions, and
\begin{equation}
\displaystyle
W(t)=\exp\left\{2\int\frac{F(t)F'(t)\left[12bF'(t)F^2(t)+c^3P^2(t)\right]}{12bF'(t)F^4(t)-6bcF'(t)P(t)+c^3F^2(t)P^2(t)}\,dt
\right\},\notag
\end{equation}

\begin{equation}
\displaystyle
P^3(t)=-\frac{108b^2}{c^6}\left[F'(t)\right]^2\left[aF'(t)-\frac{\sqrt{3}}{9}H(t)+c\right],\notag
\end{equation}

\begin{equation}
\displaystyle
H(t)=\pm\sqrt{\frac{27a^2b\left[F'(t)\right]^3+54abc\left[F'(t)\right]^2+27bc^2F'(t)-4c^3F^6(t)}{bF'(t)}}.\notag
\end{equation}

\vskip 0.5cm


\subsection {\mathversion{bold}{$\displaystyle 4a^2b^2fg^2(bcgw+ah)\left(a \frac{\partial w}{\partial
t}+cw\right)\frac{\partial^2 w}{\partial t\partial
x}=\\\\\\\left(bg\frac{\partial w}{\partial
t}-h\right)\left[4a^2b^2g^2k^2\left(\frac{\partial w}{\partial
t}\right)^2+\left(4a^3b^2cfg^2\frac{\partial w}{\partial
x}+\right.\right.\\\\\\\left.\left.4ab^2cg^2k(g+3k)w+4a^2bghk(g+k)\right)\frac{\partial
w}{\partial t}+\right. \\\\\\ \left.4a^2b^2c^2g^2fw\frac{\partial
w}{\partial x}+b^2c^2g^2(g+3k)w^2+\right. \\\\\\
\left.2abcgh(g+k)(g+3k)w+a^2h^2(g+k)^2 \right] \,,$}} where $a\neq
0$, $b\neq0$, $c$, $f\neq0$, $g\neq0$, $h$ and  $k$ are constants.

\vskip 0.5cm General solution

\begin{equation}
\displaystyle w(t,x) =-\frac{1}{bgE(t,x)}
\left[h(g+k)\int\frac{E(t,x)[V(t,x)+1]}{2kV(t,x)-g-k}\,dt+G(x)\right]\,,
\notag
\end{equation}
here

\begin{equation}
\displaystyle
E(t,x)=\exp\left[\frac{c(g+3k)}{a}\int\frac{V(t,x)}{2kV(t,x)-g-k}\,dt\right]\,,
\notag
\end{equation}

\begin{equation}
\displaystyle
V(t,x)=W\left(-\frac{g+k}{2bf}\exp\left\{-\frac{(g+3k)^2[F(t)+x]}{4a^2f}
\right\}\right) \,, \notag
\end{equation}
where $W(z)$ is the Lambert W function, $F(t)$ and $G(x)$ are
arbitrary functions. \vskip 0.5cm


\subsection {\mathversion{bold}{$\ \displaystyle V\left(\frac{\partial
w}{\partial t}+bw\right)\left(\frac{\partial^2 w}{\partial t\partial
x}+b\frac{\partial w}{\partial x}\right)+a=0\,,$}} \vskip 0.5cm

where $a, b$ are constants and $V(z)$ is any function. \vskip 0.5cm
General solution

\begin{equation}
\displaystyle w(x,t) = \left(\int W(t,x)e^{bt}\,dt+G(x)
\right)e^{-bt}\,, \notag
\end{equation}
where $W(t,x)$ is any solution of the following transcendental
equation
\begin{equation}
\displaystyle \int_s^{W(t,x)}\,V(\xi)\,d\xi +ax=F(t)\,,\notag
\end{equation}
$F(t)$ and $G(x)$ are arbitrary functions, and $s$ is an arbitrary
constant.

\vskip 0.5cm


\subsection {\mathversion{bold}{$\ \displaystyle V\left(\frac{\frac{\partial^2 w}{\partial t\partial
x}}{(2aw+b)\frac{\partial w}{\partial x}}\right)+\frac{\partial
w}{\partial t}=\frac{w(aw+b)\frac{\partial^2 w}{\partial t\partial
x}}{(2aw+b)\frac{\partial w}{\partial x}}\,,$}} \vskip 0.5cm

where $a, b$ are constants, and $V(z)$ is any function. \vskip 0.5cm
General solution

\begin{align}
\displaystyle &w(x,t) = F(t)-\exp\left\{ \int
W(t)\left[2aF(t)+b\right]dt\right\}\times\notag\\\notag\\\notag
&\left\{a\int W(t)\exp\left[ \int
W(t)\left[2aF(t)+b\right]dt\right]dt+G(x) \right\}^{-1}\,, \notag
\end{align}
where $W(t)$ is any solution of the following transcendental
equation
\begin{equation}
\displaystyle
V\left[W(t)\right]+F'(t)=W(t)F(t)\left[aF(t)+b\right]\,,\notag
\end{equation}
$F(t)$ and $G(x)$ are arbitrary functions.

\vskip 0.5cm


\subsection {\mathversion{bold}{$\displaystyle \frac{\partial^2 w}{\partial t\partial
x}+akwV\left(\frac{1}{aw}\frac{\partial w}{\partial
t}+\frac{c}{aw}+b\right)-\\\\\\\frac{1}{w}\left[\frac{\partial
w}{\partial t}+c\right]\frac{\partial w}{\partial x}=0 \,,$}} where
$a\neq 0$, $b$, $c$ are constants, and $V(z)\neq 0$ is any function.

\vskip 0.5cm General solution

\begin{align}
\displaystyle w(t,x) =&\left\{-c\int\,\exp\left[abt-a\int W(t,x)\,dt
\right]\,dt+G(x)\right\}\times\notag\\\notag\\\notag
 &\exp \left[-abt+a\int W(t,x)\,dt \right]\,,
\notag
\end{align}
here $W(t,x)$ is any solution of the following transcendental
equation

\begin{equation}
\displaystyle \int_s^{W(t,x)}\frac{d\xi}{V(\xi)}+x+F(t)=0\,, \notag
\end{equation}
where $F(t)$ and $G(x)$ are arbitrary functions, $s$ is any
constant. \vskip 0.5cm


\subsection {\mathversion{bold}{$\ \displaystyle V\left(\frac{w\frac{\partial w}{\partial x}}{\frac{\partial^2 w}{\partial t\partial
x}+b\frac{\partial w}{\partial x}}\right)+\frac{2aw\frac{\partial
w}{\partial x}\left(\frac{\partial w}{\partial t}+bw
\right)}{\frac{\partial^2 w}{\partial t\partial x}+b\frac{\partial
w}{\partial x}}=aw^2\,,$}} \vskip 0.5cm

where $a\neq 0$, $b$ are constants, and $V(z)$ is any function.
\vskip 0.5cm General solution

\begin{align}
\displaystyle &w(x,t) = F(t)-\exp\left\{
\int\frac{2aF(t)}{W(t)}dt-bt\right\}\times\notag\\\notag\\\notag
&\left\{a\int \frac{1}{W(t)}\exp\left[
\int\frac{2aF(t)}{W(t)}dt-bt\right]dt+G(x) \right\}^{-1}\,, \notag
\end{align}
where $W(t)$ is any solution of the following transcendental
equation
\begin{equation}
\displaystyle
V\left[\frac{W(t)}{2a}\right]+W(t)[F'(t)+bF(t)]=aF^2(t)\,,\notag
\end{equation}
$F(t)$ and $G(x)$ are arbitrary functions.

\vskip 0.5cm


\subsection {\mathversion{bold}{$\displaystyle
\left[(a_2b_1-a_1b_2)w-a_1b_3+b_1a_3\right]\frac{\partial^2
w}{\partial t\partial x} =\\\\\\ \left[(a_2b_1-a_1b_2)\frac{\partial
w}{\partial t}+
a_2b_3-a_3b_2\right]\,\frac{\partial w}{\partial x} \\\\\\
-(b_1\frac{\partial w}{\partial t}+b_2w+b_3)^2\,V
\left(\frac{a_1\frac{\partial w}{\partial
t}+a_2w+a_3}{b_1\frac{\partial w}{\partial t}+b_2w+b_3}\right)\,,$}}
\vskip 0.5cm

where $V\neq 0$ is an arbitrary function, $a_i$ and $b_i$ are
constants. \vskip 0.5cm General solution

\begin{align}
&w(t,x)
=\exp\left[\int\frac{-a_2+b_2Y(t,x)}{a_1-b_1Y(t,x)}\,dt\right]
\left\{\int\frac{-a_3+b_3Y(t,x)}{a_1-b_1Y(t,x)}\right.\notag\\\notag\\
&\left.\exp\left[-\int\frac{-a_2+b_2Y(t,x)}{a_1-b_1Y(t,x)}\,dt\right]\,dt+G(x)\right\}
\,, \notag
\end{align}
where the function $Y(t,x)$ is determined by the transcendental
equation
\begin{equation}
\int^{Y}_s\frac{dz}{V(z)}=x+F(t)\,\notag
\end{equation}
and $F(t)$ and $G(x)$ are arbitrary functions,  $s$ is any constant.

\vskip 1cm
\section{Equations of the Form {\mathversion{bold}{$\displaystyle \frac{\partial^2 w}{\partial
t\partial x}= F(w,\frac{\partial w}{\partial t},\frac{\partial
w}{\partial x},\frac{\partial^2 w}{\partial x^2}) $}}}

\vskip 0.5cm


\subsection {\mathversion{bold}{$\displaystyle \frac{\partial^2 w}{\partial
t \, \partial x}  = w \frac{\partial^2 w}{\partial t^2 }\,.$}}

\vskip 0.5cm General solution

\begin{equation}
\displaystyle  w(t,x) =
\int_s^{W}F(\xi)\,e^{-x/\xi}\,d\xi+G\,'(x)\,, \notag
\end{equation}
where $W=W(t,x)$ is a solution of the following transcendental
equation
\begin{equation}
\displaystyle t-\int_v^{W}\xi F(\xi)\,e^{-x/\xi}\,d\xi+G(x)=0\,
\notag
\end{equation}
and $F(\xi)$ and $G(x)$ are arbitrary functions, $s$ and $v$ are
arbitrary constants.

\vskip 0.5cm


\subsection {\mathversion{bold}{$\displaystyle (\frac{\partial w}{\partial x}+w^2)\frac{\partial^2 w}{\partial
t  \partial x}= \left(\frac{\partial^2 w}{\partial x^2
}+3w\frac{\partial w}{\partial x}+w^3\right)\frac{\partial
w}{\partial t}\,$}}

\vskip 0.5cm General solution in implicit form ($w=w(t,x) $):

\begin{equation}
\displaystyle  \int_s^{\frac{1-xw}{w}}G\,'(\xi)\xi
d\xi+xG\left(\frac{1-xw}{w}\right)= F(t)\,, \notag
\end{equation}
where  $F(t)$ and $G(z)$ are arbitrary functions, $s$ is an
arbitrary constant.

\vskip 0.5cm


\subsection {\mathversion{bold}{$\displaystyle \frac{\partial^2 w}{\partial
t  \partial x}+\frac{1}{w}\frac{\partial w}{\partial
x}=\frac{1}{\frac{\partial w}{\partial x}}\left(\frac{\partial
w}{\partial t}-1\right)\frac{\partial^2 w}{\partial x^2 }\,$}}

\vskip 0.5cm General solution

\begin{equation}
\displaystyle w(t,x)=G\,'[W(t,x)+F'(t)]+t\,, \notag
\end{equation}
where $W(t,x)$ is any solution of the following transcendental
equation
\begin{equation}
\displaystyle G[W(t,x)+F'(t)]+tW(t,x)=F(t)-tF'(t)+x\,, \notag
\end{equation}
and $F(t)$ and $G(z)$ are arbitrary functions.

\vskip 0.5cm


\subsection {\mathversion{bold}{$\displaystyle \frac{\partial^2 w}{\partial
t  \partial x}=w\frac{\partial^2 w}{\partial x^2
}+n\left(\frac{\partial w}{\partial x}\right)^2\,$}}

where $n\neq 0$ is a constant.\vskip 0.5cm General solution

\begin{equation}
\displaystyle
w(t,x)=-\frac{1}{n}\int_s^{W(t,x)}\left[G(\xi)+t\right]^\frac{1-n}{n}\,d\xi+F\,'(t)\,,
\notag
\end{equation}
where $W(t,x)$ is any solution of the following transcendental
equation

\begin{equation}
\displaystyle
\int_s^{W(t,x)}\left[G(\xi)+t\right]^\frac{1}{n}\,d\xi=x+F(t) \,,
\notag
\end{equation}
and $F(t)$ and $G(\xi)$ are arbitrary functions, $s$ is any
constant.

\vskip 0.5cm


\subsection {\mathversion{bold}{$\displaystyle \frac{\partial w}{\partial x}\frac{\partial^2 w}{\partial
t  \partial x}= \left[\frac{\partial^2 w}{\partial x^2
}+a\left(\frac{\partial w}{\partial x}\right)^2\right]\frac{\partial
w}{\partial t}\,,$}}

where $a$ is a constant.\vskip 0.5cm General solution in implicit
form ($w=w(t,x) $):

\begin{equation}
\displaystyle  e^{aw}F(t)= x+G(w)\,, \notag
\end{equation}
where  $F(t)$ and $G(z)$ are arbitrary functions.

\vskip 0.5cm


\subsection {\mathversion{bold}{$\displaystyle w\left(\frac{\partial w}{\partial x}+a\right)\frac{\partial^2 w}{\partial
t  \partial x}=\\\\\\ \left[w\frac{\partial^2 w}{\partial x^2
}-a\frac{\partial w}{\partial x}-a^2\right]\frac{\partial
w}{\partial t}\,,$}}

where $a\neq 0$ is a constant.\vskip 0.5cm General solution

\begin{equation}
\displaystyle w(t,x)=-ae^{aW(t,x)}F(t)+G\,'[W(t,x)]\,, \notag
\end{equation}
where $W(t,x)$ is any solution of the following transcendental
equation
\begin{equation}
\displaystyle G[W(t,x)]=e^{aW(t,x)}-x\,, \notag
\end{equation}
and $F(t)$ and $G(z)$ are arbitrary functions.

\vskip 0.5cm


\subsection {\mathversion{bold}{$\displaystyle \frac{\partial w}{\partial x}\frac{\partial^2 w}{\partial t^2}
-\left(\frac{\partial w}{\partial t}+a\right)\frac{\partial^2
w}{\partial t\partial x} =\\\\\\ \frac{\partial w}{\partial
x}\left[2\left(\frac{\partial w}{\partial t}\right)^2+3a
\frac{\partial w}{\partial t}+a^2 \right]\,,$}} \vskip 0.5cm

where $a\neq 0$ is a constant. \vskip 0.5cm General solution  in
implicit form ($w=w(t,x)$)

\begin{equation}
\displaystyle e^{at+2w}+F(at+w)e^{at+w}+G(x) = 0\,, \notag
\end{equation}
where $F(z)$ and $G(x)$ are arbitrary functions. \vskip 0.5cm


\subsection {\mathversion{bold}{$\displaystyle \frac{\partial w}{\partial x}\left(\frac{\partial^2 w}{\partial
t  \partial x}\right)^2+a\left(\frac{\partial^2 w}{\partial x^2
}\right)^2= \frac{\partial w}{\partial t}\frac{\partial^2
w}{\partial x^2 }\frac{\partial^2 w}{\partial t  \partial x}\,,$}}

where $a$ is a constant.\vskip 0.5cm General solution

\begin{equation}
\displaystyle w(t,x)=-a\int \frac{dt}{F'(t)}+G[x-F(t)]\,, \notag
\end{equation}
where  $F(t)$ and $G(z)$ are arbitrary functions.

\vskip 0.5cm


\subsection {\mathversion{bold}{$\displaystyle w\left(\frac{\partial w}{\partial x}-aw+1\right)\frac{\partial^2 w}{\partial
t  \partial x}=\\\\\\ \left[w\frac{\partial^2 w}{\partial x^2
}-\frac{\partial w}{\partial x}-(aw-1)^2\right]\frac{\partial
w}{\partial t}\,$}}

where $a\neq 0$ is a constant.\vskip 0.5cm General solution

\begin{equation}
\displaystyle  w(t,x)=\frac{1}{a}+e^{ax}G\,'[W(t,x)] \,, \notag
\end{equation}

where $W(t,x)$ is any solution of the following transcendental
equation
\begin{equation}
\displaystyle  W(t,x)e^{-ax}+aG[W(t,x)]=F(t) \,, \notag
\end{equation}

and  $F(t)$ and $G(z)$ are arbitrary functions.

\vskip 0.5cm


\subsection {\mathversion{bold}{$\displaystyle w\left(\frac{\partial w}{\partial x}+w\right)\frac{\partial^2 w}{\partial
t  \partial x}=\\\\\\ \left[w\frac{\partial^2 w}{\partial x^2
}+\left(\frac{\partial w}{\partial x}\right)^2+w(a+2)\frac{\partial
w}{\partial x}+aw^2\right]\frac{\partial w}{\partial t}\,$}}

where $a\neq 1$ is a constant.\vskip 0.5cm General solution in
implicit form ($w=w(t,x) $):

\begin{equation}
\displaystyle  w^{a-1}+(a-1)\left[w^ae^{ax}F(t)+G(x+\ln w)\right]=0
\,, \notag
\end{equation}
where  $F(t)$ and $G(z)$ are arbitrary functions.

\vskip 0.5cm


\subsection {\mathversion{bold}{$\displaystyle \left(\frac{\partial w}{\partial x}+a\right)\frac{\partial^2 w}{\partial
t  \partial x}=\\\\\\ \left[\frac{\partial^2 w}{\partial x^2
}-2\left(\frac{\partial w}{\partial x}\right)^2-3a\frac{\partial
w}{\partial x}-a^2\right]\frac{\partial w}{\partial t}\,$}}

where $a\neq 0$ is a constant.\vskip 0.5cm General solution in
implicit form ($w=w(t,x) $):

\begin{equation}
\displaystyle  e^w+ae^{-(w+ax)}F(t)+G\left(\frac{w}{a}+x
 \right)=0 \,, \notag
\end{equation}
where $F(t)$ and $G(z)$ are arbitrary functions.

\vskip 0.5cm


\subsection {\mathversion{bold}{$\displaystyle \left(\frac{\partial w}{\partial x}-(w+a)(2w+a)\right)\frac{\partial^2 w}{\partial
t  \partial x}=\\\\\\ \left[\frac{\partial^2 w}{\partial x^2
}-(6w+4a)\frac{\partial w}{\partial
x}+(w+a)(2w+a)^2\right]\frac{\partial w}{\partial t}\,$}}

where $a\neq 0$ is a constant.\vskip 0.5cm General solution

\begin{equation}
\displaystyle
w(t,x)=a\frac{a^2e^{-[W+ax]}F(t)-G\,'\left(\frac{W}{a}+x
 \right)}{e^{W}-a^2e^{-[W+ax]}F(t)+G\,'\left(\frac{W}{a}+x
 \right)} \,, \notag
\end{equation}
where $W=W(t,x)$ is any solution of the following transcendental
equation

\begin{equation}
\displaystyle e^{W}+ae^{-[W+ax]}F(t)+G\left(\frac{W}{a}+x \right)=0
\,, \notag
\end{equation}
and $F(t)$ and $G(z)$ are arbitrary functions.

\vskip 0.5cm


\subsection {\mathversion{bold}{$\displaystyle w\left[\frac{\partial w}{\partial x}\frac{\partial^2 w}{\partial t  \partial
x}+\frac{\partial w}{\partial t}\frac{\partial^2 w}{\partial
x^2}\right]+\frac{\partial w}{\partial t}\frac{\partial w}{\partial
x}\left[\frac{\partial w}{\partial x}+aw\right]=0\,,$}} where $a$ is
a constant. \vskip 0.5cm General solution in implicit form

\begin{equation}
\displaystyle w(t,x)^3=\frac{e^{-ax}}{aW}\left[a
e^{ax}G(W)-3W\right]\left[W+F(t)\right]\,, \notag
\end{equation}
where $W=W(t,x)$ is any solution of the following transcendental
equation
\begin{equation}
\displaystyle
ae^{ax}W\left[W+F(t)\right]G\,'(W)-3W^2-ae^{ax}F(t)G(W)=0\notag
\end{equation}
and $F(t)$ and $G(z)$ are arbitrary functions.

\vskip 0.5cm


\subsection {\mathversion{bold}{$\displaystyle \left[\frac{\partial^2 w}{\partial
x^2}+a\frac{\partial w}{\partial x}\right]\left[\frac{\partial^2
w}{\partial t  \partial x}+a\frac{\partial w}{\partial
t}\right]=b\,$}} where $a,b$ are constants. \vskip 0.5cm General
solution

\begin{equation}
\displaystyle w(t,x)=\left\{\pm \int
e^{ax}\sqrt{\frac{[2tW+G(W)][2bx+W]}{W}}\,dx+F(t)\right\}e^{-ax}\,,
\notag
\end{equation}
where $W=W(t,x)$ is any solution of the following transcendental
equation
\begin{equation}
\displaystyle W(2bx+W)G\,'(W)+2tW^2-2bxG(W)=0\notag
\end{equation}
and $F(t)$ and $G(z)$ are arbitrary functions.

\vskip 0.5cm


\subsection {\mathversion{bold}{$\displaystyle \left[\frac{\partial w}{\partial
x}+aw\right]\left[\frac{\partial^2 w}{\partial x^2}+a\frac{\partial
w}{\partial x}\right]\left[\frac{\partial^2 w}{\partial t  \partial
x}+a\frac{\partial w}{\partial t}\right]=b\,$}} where $a,b$ are
constants. \vskip 0.5cm General solution

\begin{equation}
\displaystyle w(t,x)=\left\{\frac{(144b)^{\frac{1}{3}}}{4} \int
\frac{\left[G(W)-xW-t\right]^{\frac{2}{3}}}{W^{\frac{1}{3}}}e^{ax}\,dx+F(t)\right\}e^{-ax}\,,
\notag
\end{equation}
where $W=W(t,x)$ is any solution of the following transcendental
equation
\begin{equation}
\displaystyle 2WG\,'(W)-xW+t-G(W)=0\notag
\end{equation}
and $F(t)$ and $G(z)$ are arbitrary functions.

\vskip 0.5cm


\subsection {\mathversion{bold}{$\displaystyle \frac{\partial w}{\partial x}\frac{\partial^2 w}{\partial t  \partial
x}=\left[\frac{\partial w}{\partial t}+bw\left(\frac{\partial
w}{\partial x}\right)^2\right]\frac{\partial^2 w}{\partial
x^2}+aw\left(\frac{\partial w}{\partial x}\right)^2\,,$}} where $a$,
and $b\neq 0$ are constants. \vskip 0.5cm General solution in
implicit form

\begin{equation}
\displaystyle \int_s^{w(t,x)}\,\frac{d\xi}{W(t,\xi)}=x+F(t)\,,
\notag
\end{equation}
where $W(t,\xi)=W$ is any solution of the following transcendental
equation
\begin{equation}
\displaystyle
(2a\xi+bW^2)\left[G(2a\xi+bW^2)+t\right]-\ln(-bW^2)+\ln(\xi)=0\notag
\end{equation}
and $F(t)$ and $G(z)$ are arbitrary functions, $s$ is any constant.

\vskip 0.5cm


\subsection {\mathversion{bold}{$\displaystyle \frac{\partial^2w}{\partial
t\partial x} = \left (\frac{\partial w}{\partial t}+aw\right
)\frac{\partial^2w}{\partial x^2}\left ( \frac{\partial w}{\partial
x} \right )^{-1} -a \frac{\partial w}{\partial x}+b\left (
\frac{\partial w}{\partial t}+a w\right )\,,$}} \vskip 0.5cm

where $a,b$ are constants. \vskip 0.5cm General solution

\begin{equation}
\displaystyle w(t,x) = e^{-at}G\left(F(t)+e^{-bx}\right)\,, \notag
\end{equation}
where $F(t)$ and $G(z)$ are arbitrary functions. \vskip 0.5cm


\subsection {\mathversion{bold}{$\displaystyle \frac{\partial^2w}{\partial
t\partial x} = \left (\frac{\partial w}{\partial t}+aw\right
)\frac{\partial^2w}{\partial x^2}\left ( \frac{\partial w}{\partial
x} \right )^{-1} +c \frac{\partial w}{\partial x}+b\left (
\frac{\partial w}{\partial x}+k w\right )\,,$}}  where $a$, $b$,
$c\neq-(a+b)$, and  $k$ are constants.\vskip 0.5cm

\vskip 0.5cm General solution  in implicit form

\begin{equation}
\displaystyle  \int_s^{w(t,x)}\frac{(a+b+c)\,d\xi}{G\left[\xi
e^{at}\right]e^{(b+c)t}+bk\xi}+F(t)+x =0\,, \notag
\end{equation}
where  $F(t)$ and $G(z)$ are arbitrary functions, and $s$ is an
arbitrary constant.

\vskip 0.5cm


\subsection {\mathversion{bold}{$\displaystyle \frac{\partial w}{\partial x}\frac{\partial^2 w}{\partial t  \partial
x}=\frac{\partial w}{\partial t}\frac{\partial^2 w}{\partial
x^2}+aw^m\left(\frac{\partial w}{\partial x}\right)^n\,,$}} where
$a$, $m$, and $n\neq 2$ are constants. \vskip 0.5cm General solution
in implicit form

\begin{equation}
\displaystyle \int_s^{w(t,x)}\,\frac{d\xi}{W(t,\xi)}=x+F(t)\,,
\notag
\end{equation}
where $W(t,\xi)=W$ is any solution of the following transcendental
equation
\begin{equation}
\displaystyle W^{(2-n)}+at\xi^m(n-2)=G(\xi)\notag
\end{equation}
and $F(t)$ and $G(z)$ are arbitrary functions, $s$ is any constant.

\vskip 0.5cm


\subsection {\mathversion{bold}{$\displaystyle \frac{\partial w}{\partial x}\frac{\partial^2 w}{\partial t  \partial
x}=\left[\frac{\partial w}{\partial t}+bw^n\right]\frac{\partial^2
w}{\partial x^2}+aw^m\left(\frac{\partial w}{\partial
x}\right)^2\,,$}} where $a$, $b\neq 0$, $n\neq 1$ and $m\neq n-1$
are constants. \vskip 0.5cm General solution in implicit form

\begin{equation}
\displaystyle
\int_s^{w(t,x)}\,\exp\left[\frac{a\xi^{(m-n+1)}}{b(m-n+1)}\right]\,G\left[\frac{\xi^{(1-n)}-bt(n-1)}{b(n-1)}\right]\,d\xi=F(t)-x\,,
\notag
\end{equation}
where $F(t)$ and $G(z)$ are arbitrary functions, $s$ is any
constant.

\vskip 0.5cm


\subsection {\mathversion{bold}{$\displaystyle w\left(c\frac{\partial w}{\partial x}+bw^3\right)\frac{\partial^2 w}{\partial
t  \partial x}=\\\\\\ \left[cw\frac{\partial^2 w}{\partial x^2
}-c\left(\frac{\partial w}{\partial
x}\right)^2+w(2bw^2+ac)\frac{\partial w}{\partial
x}+abw^4\right]\frac{\partial w}{\partial t}\,$}}

where $a$, $b\neq 0$, and $c$ are constants.\vskip 0.5cm General
solution in implicit form ($w=w(t,x) $):

\begin{equation}
\displaystyle  \sqrt{2}\exp\left[\frac{ac}{2bw^2}-ax
\right]\emph{erf}\left[\frac{\sqrt{2ac}}{2w\sqrt{b}}
\right]+G\left[\frac{c}{w^2}-2bx
 \right]=F(t) \,, \notag
\end{equation}
where $\emph{erf}(z)$ is the error function, $F(t)$ and $G(z)$ are
arbitrary functions.

\vskip 0.5cm


\subsection {\mathversion{bold}{$\displaystyle \frac{\partial w}{\partial x}\frac{\partial^2 w}{\partial t  \partial
x}-\frac{\partial w}{\partial t}\frac{\partial^2 w}{\partial
x^2}+[bV(w)+a]\left(\frac{\partial w}{\partial x}\right)^2+\\\\\\
\frac{V(w)[bV(w)+a]}{V\,'(w)}\frac{\partial^2 w}{
\partial x^2}=0\,,$}} where $a\neq 0$, and $b$ are constants, $V(w)\neq const$ is any function.
\vskip 0.5cm General solution in implicit form

\begin{equation}
\displaystyle
\int_s^{w(t,x)}\,V(\xi)G\left\{\frac{e^{at}\left[bV(\xi)+a\right]}{aV(\xi)}\right\}\,d\xi+x+F(t)=0\,,
\notag
\end{equation}
where  $F(t)$ and $G(z)$ are arbitrary functions, $s$ is any
constant.

\vskip 1cm
\section{Equations of the Form \\\\{\mathversion{bold}{$\displaystyle \frac{\partial^2 w}{\partial
t\partial x}= F(w,\frac{\partial w}{\partial t},\frac{\partial
w}{\partial x},\frac{\partial^2 w}{\partial t^2},\frac{\partial^2
w}{\partial x^2}) $}}}

\vskip 0.5cm


\subsection {\mathversion{bold}{$\displaystyle \left(\frac{\partial
w}{\partial x}\right)^2\frac{\partial^2 w}{\partial
t^2}-\left(\frac{\partial w}{\partial t}\right)^2\frac{\partial^2
w}{\partial x^2}=0\,.$}} \vskip 0.5cm General solution

\begin{equation}
\displaystyle
w(t,x)=F\left[\frac{\left(xW-G(W)+t\right)^2}{W}\right]\,, \notag
\end{equation}
where $W=W(t,x)$ is any solution of the following transcendental
equation
\begin{equation}
\displaystyle G(W)-2WG\,'(W)+xW-t=0\notag
\end{equation}
and $F(z)$ and $G(z)$ are arbitrary functions.

\vskip 0.5cm


\subsection {\mathversion{bold}{$\displaystyle 2\frac{\partial w}{\partial t}\frac{\partial w}{\partial x}\frac{\partial^2 w}{\partial t  \partial
x}- \left(\frac{\partial w}{\partial x}\right)^2\frac{\partial^2
w}{\partial t^2}-\left(\frac{\partial w}{\partial
t}\right)^2\frac{\partial^2 w}{\partial x^2}=0\,.$}} \vskip 0.5cm
General solution in implicit form ($w=w(t,x) $):

\begin{equation}
\displaystyle G(w)\left[F(w)+x\right]+t=0\,, \notag
\end{equation}
where $F(z)$ and $G(z)$ are arbitrary functions.

\vskip 0.5cm


\subsection {\mathversion{bold}{$\displaystyle \left[a\frac{\partial w}{\partial t}+2\frac{\partial w}{\partial t}\frac{\partial w}{\partial x}+b\frac{\partial w}{\partial x}\right]\frac{\partial^2 w}{\partial t  \partial
x}=\\\\\\\frac{\partial w}{\partial x}\left(a+\frac{\partial
w}{\partial x}\right)\frac{\partial^2 w}{\partial
t^2}+\frac{\partial w}{\partial t}\left(b+\frac{\partial w}{\partial
t}\right)\frac{\partial^2 w}{\partial x^2}\,,$}}where $a\neq 0$, and
$b$ are constants. \vskip 0.5cm General solution in implicit form
($w=w(t,x) $):

\begin{equation}
\displaystyle
a\int_s^{\frac{bt+ax}{a}}\frac{d\xi}{aG(w+a\xi)-b}+F(w)+t=0\,,
\notag
\end{equation}
where $F(z)$ and $G(z)$ are arbitrary functions, $s$ is any
constant.

\vskip 0.5cm


\subsection {\mathversion{bold}{$\displaystyle \left(\frac{\partial^2 w}{\partial t  \partial
x}\right)^2-\frac{\partial^2 w}{\partial t^2}\frac{\partial^2
w}{\partial x^2}+a\frac{\partial^2 w}{\partial t^2}=0\,,$}} where
$a$ is a constant. \vskip 0.5cm General solution

\begin{equation}
\displaystyle w(t,x)=\frac{at^2}{2}+tG(W)+F(W)+xW\,, \notag
\end{equation}
where $W=W(t,x)$ is any solution of the following transcendental
equation
\begin{equation}
\displaystyle tG\,'(W)+F\,'(W)+x=0\notag
\end{equation}
and $F(z)$ and $G(z)$ are arbitrary functions.

\vskip 0.5cm


\subsection {\mathversion{bold}{$\displaystyle \left(\frac{\partial^2 w}{\partial t  \partial
x}\right)^2-\frac{\partial^2 w}{\partial t^2}\frac{\partial^2
w}{\partial x^2}+a\frac{\partial^2 w}{\partial t^2}\frac{\partial
w}{\partial x}=0\,,$}} where $a\neq 0$ is a constant. \vskip 0.5cm
General solution

\begin{equation}
\displaystyle w(t,x)=\frac{1}{a}\left[atW+aG(W)+F(W)e^{ax}\right]\,,
\notag
\end{equation}
where $W=W(t,x)$ is any solution of the following transcendental
equation
\begin{equation}
\displaystyle aG\,'(W)+e^{ax}F\,'(W)+at=0\notag
\end{equation}
and $F(z)$ and $G(z)$ are arbitrary functions.

\vskip 0.5cm


\subsection {\mathversion{bold}{$\displaystyle \left(\frac{\partial^2 w}{\partial t  \partial
x}\right)^2-\frac{\partial^2 w}{\partial t^2}\frac{\partial^2
w}{\partial x^2}+a\left(\frac{\partial w}{\partial
t}\right)^2\frac{\partial^2 w}{\partial x^2}=0\,,$}} where $a\neq 0$
is a constant. \vskip 0.5cm General solution

\begin{equation}
\displaystyle
w(t,x)=\frac{1}{a}\left\{aF(W)+axW-\ln\left[G(W)+at\right]\right\}\,,
\notag
\end{equation}
where $W=W(t,x)$ is any solution of the following transcendental
equation
\begin{equation}
\displaystyle -G\,'(W)+axG(W)+aF\,'(W)(G(W)+at)+a^2tx=0\notag
\end{equation}
and $F(z)$ and $G(z)$ are arbitrary functions.

\vskip 0.5cm


\subsection {\mathversion{bold}{$\displaystyle \left(\frac{\partial^2 w}{\partial t  \partial
x}\right)^2+a\frac{\partial w}{\partial x}\frac{\partial^2
w}{\partial t  \partial x}+\left(b\frac{\partial w}{\partial
x}-\frac{\partial^2 w}{\partial x^2}\right)\frac{\partial^2
w}{\partial t^2}=0\,,$}} where $a$, and $b\neq 0$ are constants.
\vskip 0.5cm General solution

\begin{equation}
\displaystyle w(t,x)=\frac{Te^{bx}}{b}+F(T)+\int_s^tWd\xi\,, \notag
\end{equation}
where $W=W(T,\xi)$ is any solution of the transcendental equation
\begin{equation}
\displaystyle G(W)e^{-a\xi}+T=0\,,\notag
\end{equation}
$T=T(t,x)$ is any solution of the following transcendental equation
\begin{equation}
\displaystyle b\int_s^t\frac{e^{a\xi}d\xi}
{G\,'\left[W(T,\xi)\right]}=bF\,'(T)+e^{bx}\notag
\end{equation}
and $F(z)$ and $G(z)$ are arbitrary functions, $s$ is any constant.

\vskip 0.5cm


\subsection {\mathversion{bold}{$\displaystyle \frac{\partial w}{\partial x}\left(\frac{\partial^2 w}{\partial t  \partial
x}\right)^2+a\frac{\partial^2 w}{\partial t  \partial
x}+\frac{\partial w}{\partial x}\left(b-\frac{\partial^2 w}{\partial
x^2}\right)\frac{\partial^2 w}{\partial t^2}=0\,,$}} where $a$, and
$b\neq 0$ are constants. \vskip 0.5cm General solution

\begin{equation}
\displaystyle
w(t,x)=\pm\frac{\left(T+2bx\right)^{\frac{3}{2}}}{3b}+F(T)+\int_s^tWd\xi\,,
\notag
\end{equation}
where $W=W(T,\xi)$ is any solution of the transcendental equation
\begin{equation}
\displaystyle G(W)+2a\xi+T=0\,,\notag
\end{equation}
$T=T(t,x)$ is any solution of the following transcendental equation
\begin{equation}
\displaystyle 2b\int_s^t\frac{d\xi}
{G\,'\left[W(T,\xi)\right]}=2bF\,'(T)\pm\sqrt{2bx+T}\notag
\end{equation}
and $F(z)$ and $G(z)$ are arbitrary functions, $s$ is any constant.

\vskip 0.5cm


\subsection {\mathversion{bold}{$\displaystyle \left(\frac{\partial^2 w}{\partial t  \partial
x}\right)^2+a\left(\frac{\partial w}{\partial
x}\right)^2\frac{\partial^2 w}{\partial t  \partial
x}+\\\\\\\left[b\left(\frac{\partial w}{\partial
x}\right)^2-\frac{\partial^2 w}{\partial x^2}\right]\frac{\partial^2
w}{\partial t^2}=0\,,$}} where $a$, and $b\neq 0$ are constants.
\vskip 0.5cm General solution

\begin{equation}
\displaystyle w(t,x)=-\frac{\ln(bx-T)}{b}+F(T)+\int_s^tWd\xi\,,
\notag
\end{equation}
where $W=W(T,\xi)$ is any solution of the transcendental equation
\begin{equation}
\displaystyle G(W)-a\xi+T=0\,,\notag
\end{equation}
$T=T(t,x)$ is any solution of the following transcendental equation
\begin{equation}
\displaystyle \int_s^t\frac{d\xi}
{G\,'\left[W(T,\xi)\right]}-F\,'(T)=\frac{1}{b(bx-T)}\notag
\end{equation}
and $F(z)$ and $G(z)$ are arbitrary functions, $s$ is any constant.

\vskip 0.5cm


\subsection {\mathversion{bold}{$\displaystyle \left(\frac{\partial^2 w}{\partial t  \partial
x}\right)^2+a\frac{\partial^2 w}{\partial t  \partial
x}-\frac{\partial^2 w}{\partial t^2}\frac{\partial^2 w}{\partial
x^2}+b\frac{\partial^2 w}{\partial t^2}+c\frac{\partial^2
w}{\partial x^2}=bc\,,$}} where $a$, $b$, and $c$ are constants.
\vskip 0.5cm General solution

\begin{equation}
\displaystyle
w(t,x)=\frac{ct^2+bx^2}{2}+xT+F(T)+\int_s^t\,W\,d\tau\,, \notag
\end{equation}
where $W=W(T,\tau)$ is any solution of the following transcendental
equation
\begin{equation}
\displaystyle G(W)+T+a\tau=0\notag
\end{equation}
and $T=T(t,x)$ is any solution of the following transcendental
equation
\begin{equation}
\displaystyle
\int_s^t\,\frac{d\tau}{G\,'[W(T,\tau)]}=F\,'(T)+x\notag
\end{equation}
where $F(z)$ and $G(z)$ are arbitrary functions, $s$ is any
constant.

\vskip 0.5cm


\subsection {\mathversion{bold}{$\displaystyle \left[\frac{\partial^2 w}{\partial t^2}\frac{\partial^2 w}{\partial
x^2}-\left(\frac{\partial^2 w}{\partial t  \partial
x}\right)^2\right]V\,' \left(\frac{\partial w}{\partial
x}\right)=\\\\\\a\frac{\partial^2 w}{\partial t  \partial
x}-b\frac{\partial^2 w}{\partial t^2}\,,$}} where $a$, $b$ are
constants and $V(z)\neq const$ is an arbitrary function. \vskip
0.5cm General solution

\begin{equation}
\displaystyle w(t,x)=\int_s^x\,W\,d\xi+\int_v^t\,H\,d\tau+F(T) \,,
\notag
\end{equation}
where $W=W(T,\xi)$ and $H=H(T,\tau)$ are any solutions of the
following transcendental equations
\begin{equation}
\displaystyle V(W)-T+b\xi=0\,,\notag
\end{equation}
\begin{equation}
\displaystyle G(H)+T+a\tau=0\notag
\end{equation}
and $T=T(t,x)$ is any solution of the following transcendental
equation
\begin{equation}
\displaystyle
\int_s^x\,\frac{d\xi}{V\,'[W(T,\xi)]}-\int_v^t\,\frac{d\tau}{G\,'[H(T,\tau)]}+F\,'(T)\,,\notag
\end{equation}
here $F(z)$ and $G(z)$ are arbitrary functions, $s,v$ are any
constants.

\vskip 0.5cm


\subsection {\mathversion{bold}{$\displaystyle \left(\frac{\partial^2 w}{\partial t  \partial
x}\right)^2-\frac{\partial^2 w}{\partial t^2}\frac{\partial^2
w}{\partial x^2}+b\frac{\partial^2 w}{\partial t^2}+a\frac{\partial
w}{\partial x}=ab\,,$}} where $a,b$ are constants. \vskip 0.5cm
General solution

\begin{equation}
\displaystyle w(t,x)=\frac{at^2+bx^2}{2}-tG(W)+xW+F(W)\,, \notag
\end{equation}
where $W=W(t,x)$ is any solution of the following transcendental
equation
\begin{equation}
\displaystyle tG\,'(W)-F\,'(W)=x\notag
\end{equation}
and $F(z)$ and $G(z)$ are arbitrary functions.

\section{Appendix. \\\emph{Maple} procedure \emph{reduce\_PDE\_order}}
reduce\_PDE\_order:=proc(pde,unk) \\
local
B,W,N,NN,ARG,acargs,i,M,pde0,DN,IND,IND2,IND3,IND4,ARGS,SUB,SUB0,
Z0,Bargs,EQS,XXX,WW,BB,PP,pdeI,IV,s,AN,NA;

 option `Copyright (c) 2006-2007 by Yuri N. Kosovtsov. All rights reserved.`;
 \\N:=PDETools[difforder](op(1,[selectremove(has,indets(pde,function),unk)]));
 \\NN:=op(1,[selectremove(has,op(1,[selectremove(has,indets(pde,function),unk)]),diff)]);
 ARG:=[op(unk)];
 \\acargs:=\{\};
 \\for i from 1 to nops(ARG) do
 \\if PDETools[difforder](NN,op(i,ARG))=0 then else acargs:=acargs union \{op(i,ARG)\} fi; od;
 \\acargs:=convert(acargs,list);
 \\M:=op(0,unk)(op(acargs));
 \\if type(pde,equation)=true then
 \\pde0:=lhs(subs(unk=M,pde))-rhs(subs(unk=M,pde)) else pde0:=subs(unk=M,pde) fi;
 \\DN:=[seq(seq(i,i=1..nops(acargs)),j=1..N)];
 \\IND:=seq(op(combinat[choose](DN,i)),i=1..N);
 \\IND2:=seq(op(combinat[choose](DN,i)),i=1..N-2);
 \\IND3:=op(combinat[choose](DN,N-1));
 \\IND4:=op(combinat[choose](DN,N));
 \\ARGS:=op(unk),M,seq(convert(D[op(op(i,[IND2]))](op(0,unk))\\(op(acargs)),diff),i=1..nops([IND2]));
 \\SUB:=\{M=W[0],seq(convert(D[op(op(i,[IND]))](op(0,unk))\\(op(acargs)),diff)=W[op(op(i,[IND]))],i=1..nops([IND]))\};
 \\SUB0:=\{W[0]=op(0,unk)(op(ARG)),\\seq(W[op(op(i,[IND]))]=subs(M=op(0,unk)(op(ARG)),\\convert(D[op(op(i,[IND]))](op(0,unk))(op(acargs)),diff)),i=1..nops([IND]))\};
 \\Z0:=B(ARGS,seq(convert(D[op(op(i,[IND3]))](op(0,unk))(op(acargs)),diff),\\i=1..nops([IND3])));
 \\Bargs:=op(indets(subs(SUB,Z0),name));
 \\EQS:=convert(subs(SUB,\{seq(diff(Z0,op(i,acargs))=0,i=1..nops(acargs))\}),diff);
 \\XXX:=\{seq(W[op(op(i,[IND4]))],i=1..nops([IND4]))\};
 \\WW:=select(type,indets(subs(SUB,pde0)), 'name') intersect\\ \{seq(W[op(op(i,[IND4]))],i=1..nops([IND4]))\};
 \\BB:=select(has,combinat[choose](XXX, nops(acargs)),WW);
 \\PP:=\{\}; NA:=0;
 \\pdeI:=\{seq(\{subs(subs(solve(EQS,op(i,BB)),subs(SUB,pde0)))\},i=1..nops(BB))\};
 \\IV:=\{seq(W[op(op(i,[IND4]))],i=1..nops([IND4]))\};
 \\for s from 1 to nops(pdeI) do
 \\try
 \\AN:=[pdsolve(op(s,pdeI),\{B\},ivars=IV)];
 \\if AN=[NULL] then else NA:=1 fi;
 \\for i from 1 to nops(AN) do
 \\if op(0,lhs(op(i,AN)))=B then
 \\PP:=PP union \{rhs(op(i,AN))\}
 \\fi;
 \\od;
 \\catch:
 \\end try;
 \\od;
 \\PP:=subs(SUB0,PP);
 \\if NA=1 then if PP={} then print(WARNING("SOLUTION EXISTS")) fi;fi;
 \\RETURN(PP);
 \\end proc:

\vskip 1cm \textbf{Calling Sequence}:
\emph{reduce\_PDE\_order}(\textbf{PDE},\,\textbf{$f(\vec{x})$});
\medskip

\textbf{PDE}\,\,-\,\,partial differential equation;

\textbf{$f(\vec{x})$}\,\,-\,\,indeterminate function with its
arguments.

\medskip
\textbf{Notice:} The reduced PDE is $B=0$.

\end{document}